# Diffraction of Ultrashort Pulse on a Nanoscale Conductive Cone

P.A. Golovinski [1,2], V.A. Astapenko [1], E.S. Manuylovich [1]

[1] *Moscow Institute of Physics and Technology, Russia*, [2] *Physics Research Laboratory, Voronezh State University of Architecture and Civil Engineering, Russia*

*e-mail: golovinski@bk.ru*

Surface plasmon polariton is collective oscillation of the free electrons at metal dielectric interface. As a wave phenomenon, surface plasmon polaritons can be focused using appropriate excitation geometry of metallic structures. We theoretically demonstrate the possibility of controlling nanoscale short pulse superfocusing based on the generation of radially polarized surface plasmon polariton mode of conical metallic tip. Numerical simulation for femtosecond pulse propagation along a silver nano-needle is discussed. The spatial distribution for a near field strongly depends on a linear chirp of the laser pulse which can partially compensate the wave dispersion. Field distribution is calculated for different chirp values, opening angles and distances. For a pulse with a negative chirp, pulse duration becomes shorter with amplification ~40.

## 1. Introduction

One of the major problem in nano-optics is related to the concentration and delivery of electromagnetic field to the nanoscale regions [1]. This is especially important for efficient coupling of light into and out of nano-optical devices, electromagnetic probing of either separate molecules as quantum dots.

Evanescent fields play a central role in nano-optics [2]. They are characterized by the fact that the strength of the evanescent electromagnetic field decays exponentially as one moves away from the interface into either medium. In the condensed matter literature these waves commonly reffered to as surface plasmon polaritons. Shortly speaking a surface plasmon polariton is an admixture of the electromagnetic field and collective particle excitation that propagates along the surface of a medium, or along the interface between two media. In the context of near-field optics and nano-optics the study of optical phenomena related to surface waves has been termed plasmonics or nanoplasmonics [3].

In order to achieve the goal of nanometer confined femtosecond source with spatial localization and temporal control of the optical field the use of unique properties of surface plasmon polaritons (SPP) has been widely discussed as a potential tool.

Electromagnetic surface waves associated with collective charge density oscillation at metal-dielectric interfaces allow for subwavelength spatial control of even broadband optical fields [4]. One of the most prominent approach to the space-temporal electromagnetic field superfocusing is related to using plasmon propagation in sharply tapered metal nanorods [5].

The divergence of the effective index of refraction with decreasing of cone radius experienced by SPP propagating toward the apex leads to continuous transformation of cylindrical modes and to the near adiabatic SPP nanofocusing into the apex of the tip. A 3D taped tip, as an SPP waveguide, stands out due to its unique topology as a cone. It has been demonstrated experimentally, this geometry allows for true 3D focusing into excitation volume as small as a few tens of nanometers in size [6]. In contrast to other nanofocusing structures based on localized plasmon resonances that exhibit a dependence of spatial localization on the spectral and phase characteristics the external field, one can control the pulse duration and optical waveform at the tip apex via deterministic pulse shaping. The theoretical limit for the shortest attainable pulse duration at the tip apex is determined only by the coupling bandwidth and dephasing time.

The details of the nanofocusing mechanism are not yet completely understood even under adiabatic conditions, ignoring radiation damping and reflection. This new optical antenna allows to achieve far- to near-field transformation of light from micro- to the nanoscale. The aim of this paper is numerical simulation of the control super focusing via chirped pulse, based on asymptotic solution of the Maxwell equations near tip apex.

## 2. Theoretical model

Investigation of field enhancement near cone tip leads to conclusion that the process is typical for symmetric TM (m=0) waves. This type of polarization can be obtained, for example, by direct focusing of laser radiation with appropriate polarization on the base of cone [7]. In the following consideration we imply the wavepacket propagation. The main difference of this problem from the classical diffraction [8] is explained by small diameter of the nanowire compared to the depth of skin-effect and frequency dependence of complex dielectric constants for metals at the optical wavelengths.

It has been shown using 3D finite difference time domain analysis that it is possible to selectively confine either the electric or magnetic field at the tip apex [9]. We follow analytical approach [10] to the solution of the Maxwell equations. It is convenient to choose spherical system of coordinates as shown in Fig. 1 to consider the field near cone apex.

We assume the magnetic field is directed along $\varphi$ axis and depends only on the coordinates $r$ and $\theta$.

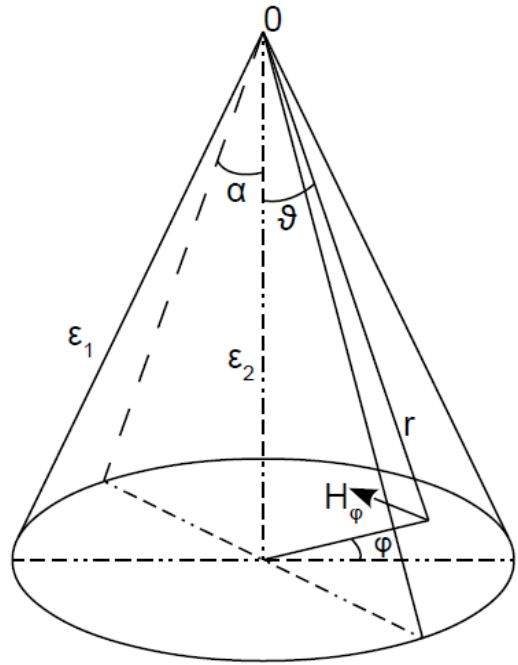

**Fig.1.** Geometry of conical structure for superfocusing.

For radially polarized modes and conical wave-guide, there is no azimuthal angle dependence and the eigenmodes satisfy the wave equation as follows [11]:

$$\frac{1}{r^2}\frac{\partial^2}{\partial r^2}(rH_\varphi) + \frac{1}{r^2}\frac{\partial}{\partial \theta}\left(\frac{1}{\sin\theta}\frac{\partial(\sin\theta H_\varphi)}{\partial \theta}\right) = \frac{\varepsilon}{c^2}\frac{\partial^2 H_\varphi}{\partial t^2}. \tag{1}$$

The boundary condition is trivial ($H_\varphi \to 0$) far from the cone surface, i.e. the field tends to zero, because plasmon polaritons are responsible for near field part of the field.

The intermodal coupling of different modes was discussed in [12]. The coupling coefficients are proportional to the taper angle $\alpha$. The effective coupling between guided modes is small up to the angles $35°$, and adiabatic approximation is applicable [13]. Adiabatic requires that the thinning of the wire is not abrupt and occurs slowly over the effective wavelengths [6]. In our consideration we neglected the coupling of modes with different m that is valid at the vicinity of the apex.

To separate variables we write down the field in the form

$$H_\varphi(r,\theta,t) = R(r)\Psi(\theta)\exp(-i\omega t), \tag{2}$$

where $\omega$ is wave frequency, $c$ is velocity of light, $\varepsilon$ is dielectric constant ($\mu = 1$). Then for $\Psi(\theta)$ and $R(r)$ we have

$$\frac{d^2\Psi}{d\theta^2} + \frac{\cos\theta}{\sin\theta}\frac{d\Psi}{d\theta} - \left(\eta^2 + \frac{1}{\sin^2\theta}\right)\Psi = 0, \tag{3}$$

$$\frac{d^2R}{dr^2} + \frac{2}{r}\frac{dR}{dr} + \left(\frac{\eta^2}{r^2} + \varepsilon\frac{\omega^2}{c^2}\right)R = 0. \tag{4}$$

Here, $\eta$ is separation constant to be determined from the boundary conditions on the surface. For small apex angle $\alpha$ the parameter $\theta \ll 1$, both inside cone and outside it near surface. Then function $\Psi$ obeys the approximate equation

$$x^2\frac{d^2\Psi}{dx^2} + x\frac{d\Psi}{d\theta} - (x^2 + 1)\Psi = 0, \tag{5}$$

where $x = \eta\theta$.

The TM$_0$ solution to the Eq. (5) can be expressed in terms of the modified Bessel and Hankel functions [14] as

$$\Psi(x) = AI_1(x) + BK_1(x). \tag{6}$$

This choice of functions is justified by evanescent nature such type a wave. For two different regions of angles, the solution we write down in the form

$$\Psi(\theta) = AI_1(\eta\theta), \theta \leq \alpha, \tag{8}$$

$$\Psi(\theta) = BK_1(\eta\theta), \theta \geq \alpha. \tag{9}$$

Let us denote $k = \sqrt{\varepsilon}\omega/c$ and introduce $R(r) = Z(\rho)/\sqrt{r}$, $\rho = kr$. After that, we will have equation

$$\rho^2 Z'' + \rho Z' + (\rho^2 + v^2) = 0, v^2 = \eta^2 - 1/4. \tag{10}$$

The Eq. (10) has solution in the form of the Bessel functions with imaginary indexes [15], having the series expansion:

$$\text{Sf}_\nu(\rho) = \left[1 - \frac{1}{1+\nu^2}\left(\frac{\rho}{2}\right)^2 + \ldots\right]\sin(\nu \ln \rho) + \left[\frac{\nu}{1+\nu^2}\left(\frac{\rho}{2}\right)^2 + \cdots\right]\cos(\nu \ln \rho), \quad (11)$$

$$\text{Cf}_\nu(\rho) = \left[1 - \frac{1}{1+\nu^2}\left(\frac{\rho}{2}\right)^2 + \ldots\right]\cos(\nu \ln \rho) + \left[\frac{\nu}{1+\nu^2}\left(\frac{\rho}{2}\right)^2 + \cdots\right]\sin(\nu \ln \rho). \quad (12)$$

For small values of $\rho \to 0$, we get asymptotically $\text{Sf}_\nu(\rho) \sim \sin(\nu \ln \rho)$, $\text{Cf}_\nu(\rho) \sim \cos(\nu \ln \rho)$.

Near a cone apex, $\eta^2/r^2 \gg |\varepsilon|\omega^2/c^2$, an asymptotic solution comes to the form

$$H_\varphi \sim \frac{A(\omega)}{\sqrt{r}}\Psi(\theta)\exp\left(-i\omega t - i\nu \ln \frac{r}{r_0}\right), \quad (13)$$

where distance $r_0$ from the apex fixes the wave phase.

Following the Maxwell equations we have

$$E_r = \frac{ic}{\omega\varepsilon}\frac{1}{r\sin\theta}\frac{\partial(H_\varphi \sin\theta)}{\partial\theta}, \quad (14)$$

$$E_\theta = -\frac{ic}{\omega\varepsilon}\frac{\partial}{\partial r}(rH_\varphi). \quad (15)$$

The final asymptotic form for electric field components is

$$E_r(\omega) = -\frac{ic}{\omega\varepsilon_2}\frac{\nu A(\omega)}{r^{3/2}}I_0(\eta\theta)\exp\left(-i\nu \ln \frac{r}{r_0}\right), \quad (16)$$

$$E_\theta(\omega) = \frac{ic}{\omega\varepsilon_2}\left(\nu + \frac{i}{2}\right)\frac{A(\omega)}{r^{3/2}}I_1(\eta\theta)\exp\left(-i\nu \ln \frac{r}{r_0}\right) \quad (17)$$

for $\theta \leq \alpha$, and

$$E_r(\omega) = \frac{ic}{\omega\varepsilon_1}\frac{\nu B(\omega)}{r^{3/2}}K_0(\eta\theta)\exp\left(-i\nu \ln \frac{r}{r_0}\right), \quad (18)$$

$$E_\theta(\omega) = \frac{ic}{\omega\varepsilon_1}\left(\nu + \frac{i}{2}\right)\frac{B(\omega)}{r^{3/2}}K_1(\eta\theta)\exp\left(-i\nu \ln \frac{r}{r_0}\right) \quad (19)$$

for $\theta \geq \alpha$. Both these solutions demonstrate anomalous growth of the wave field near the tip of the cone.

From the continuity conditions of tangential components of the field on the interface

$$\varepsilon_2\frac{I_1(\alpha\eta)}{I_0(\alpha\eta)} = -\varepsilon_1\frac{K_1(\alpha\eta)}{K_0(\alpha\eta)}. \quad (19)$$

Eq. (19) determines the egenvalue $\eta$ uniquely. As for the modified cylindrical functions, The nonequalities hold true when $I_0(x) > I_1(x)$ and $K_0(x) > K_1(x)$, that is the surface wave under investigation can exist if $|\varepsilon_2| > |\varepsilon_2|$.

The absorption effect can be taken into account when the dielectric permittivity has the complex-valued form $\varepsilon = \varepsilon' + i\varepsilon''$. Then, the parameter $\eta$ also becomes complex: $\eta = \eta' + i\eta''$.

5Because of complexity of $\eta$ there arises an additional factor $r^{\eta''}$ in the field dependence. The details of asymptotic analyses of dispersion relation Eq. (19) one can find in [10].

### 3. Plasmon polariton pulse focusing

We consider a short pulse of plasmon polaritons, propagating in a metallic cone, for initial condition

$$H_\varphi(R,\theta,t) = f_0 \exp\left(-at^2 - i\omega_0 t - i\beta t^2\right). \quad (20)$$

The Fourier transform of this pulse gives [16]

$$H_\varphi(R,\theta,\omega) = \frac{f_0}{\sqrt{4\pi(a+i\beta)}} \exp\left(-\frac{(\omega-\omega_0)^2}{4(a+i\beta)^2}\right). \quad (21)$$

Space-time evolution of the wave packet can be simply expressed as the Fourier integral

$$E_j(r,\theta,t) = \int_{-\infty}^{\infty} E_j(r,\theta,\omega) E_j(\omega) e^{-i\omega t} d\omega, \quad j = r,\theta. \quad (22)$$

For practical numerical simulations of integral in the Eq. (22) we use formerly calculated dependence of $\eta$ as a function of frequency $\omega$. The real and imaginary parts of $\eta$ for silver needle with apex angle $\alpha = 0.1$ are plotted in Fig. 2.

This result allows for obtain phase and group velocity [[17] as shown in Fig. 3. For a distance 400 nm from the apex and light wavelength $\lambda = 630$ nm the group velocity is equal to $0.35c$, and near the tip it tends to zero. This effect one can partially compensate by decreasing laser frequency during the pulse.

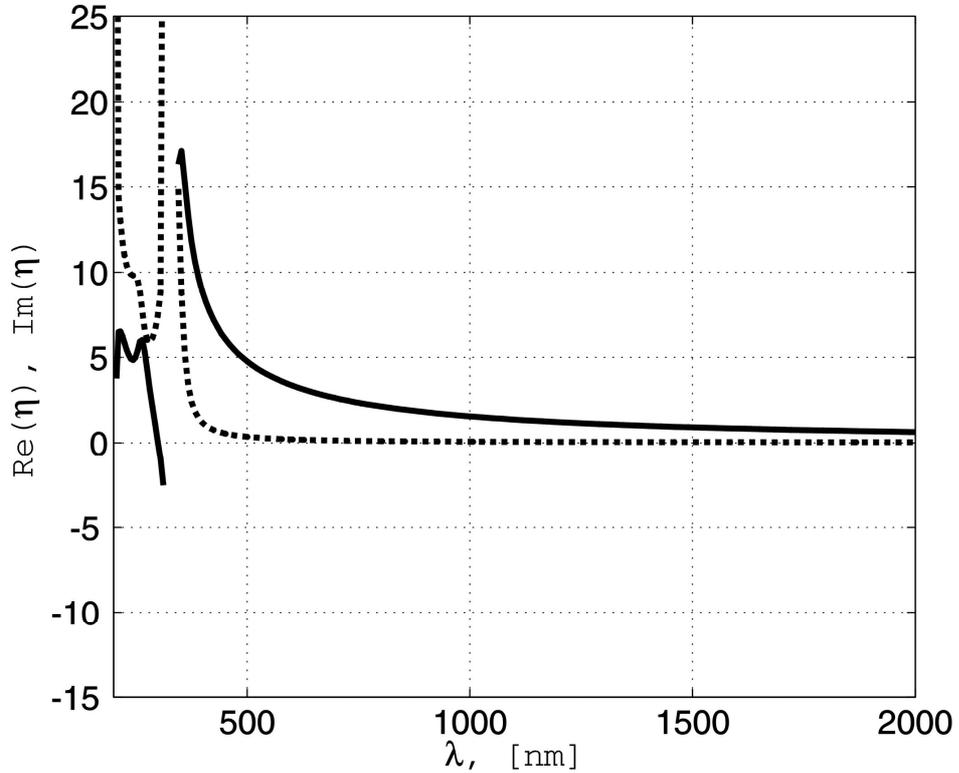

**Fig. 2.** Dependence of Re$\eta$ (solid line) and Im$\eta$ (dotted line) as a function of wavelength $\lambda$.

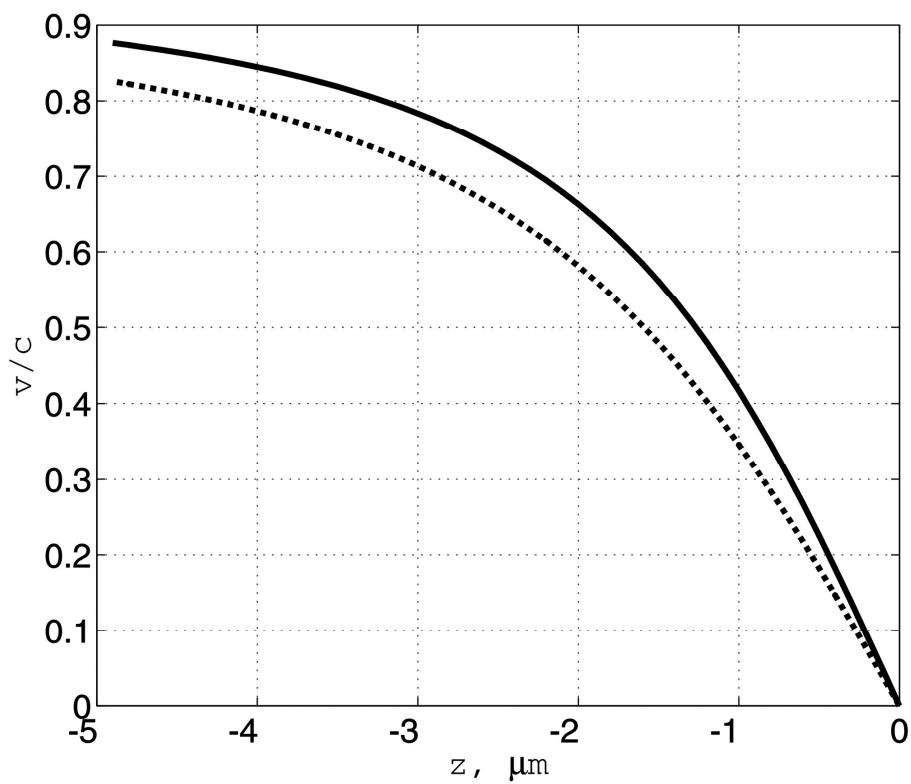

**Fig. 3.** Phase velocity (solid line) and the group velocity (dotted line) as a function of distance from the tip.

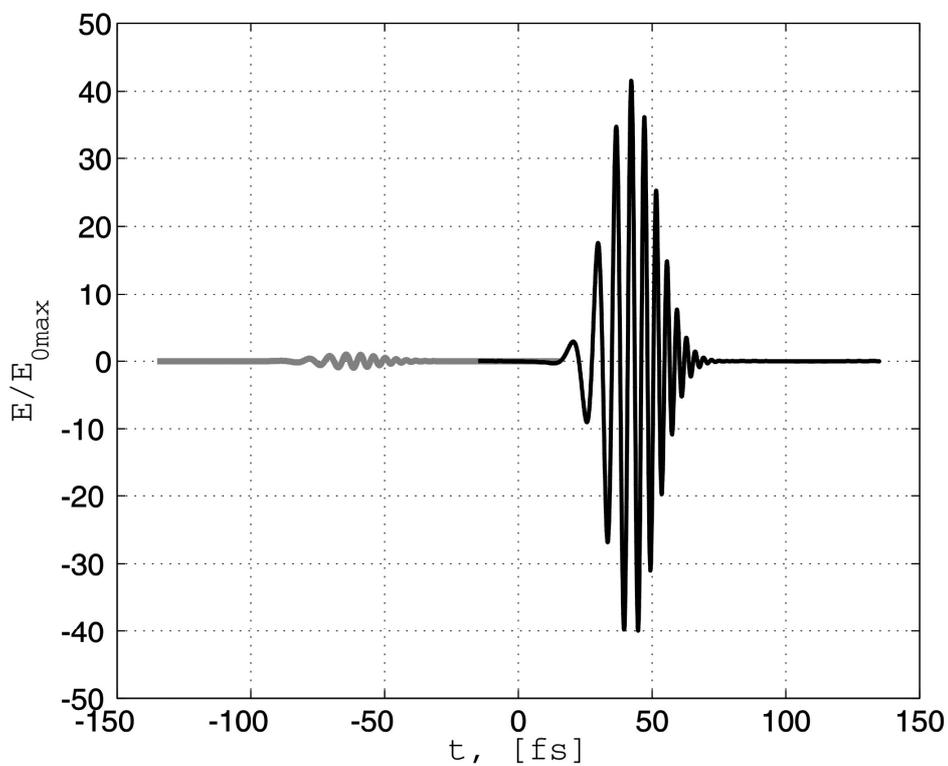

Fig. 4. Compression of plasmon polariton wave as a result of propagation starting at the distance $r_i = 1000$ nm from the apex up to the distance $r_f = 250$ nm.

Fig. 4 shows the results of numerical simulations for the field at the vicinity of silver nano needle for laser pulse with duration 32 fs, carrier wavelength 1550 nm and negative chirp $\beta$ = -0.013 fs$^{-1}$. Amplification of electric field for $\beta = 0$ fs$^{-1}$ is equal 40.2, for $\beta = 0.013$ fs$^{-1}$ is 42.7, and for $\beta = 0.013$ fs$^{-1}$ is 36.5. At the same conditions variation of pulse duration is 0 fs, -4.5 fs and 6 fs correspondingly.

## 4. Conclusions

The continuous SPP mode transformation taking advantage of radius-dependent index of refraction experienced by SPPs propagating on the outside of conical tip leads to nanofocusing at the apex. It is crucial to have a large component of excitation along the axial direction in order to obtain a hight field enhancement. The gigantic wave field can be produced in surface plasmon polariton propagation only under specific conditions. The details of design and fabrication a plasmonic structure aimed at efficient launching of radially polarized incident illumination onto needle and subsequently to the needle tip are presented in [18]. It is instructive to point out that the geometry of the cone is more suitable for superfocusing than the wedge. In the case of the cone, the wave field increases more rapidly because of more rigid space confinement. The role of reflection and radiation losses will be investigated in future research.

The ability to generate a nanoconfined optical excitation at the end of a sharp probe tip with high nanofocusing efficiency holds significant promise for near-field control of nanostructures. Extension of ultrafast pulse manipulation to the nanoscale through plasmonic nanofocusing will allow for the all-optical control of the elementary excitation of matter on their characteristic time and length scales simultaneously. For a first order process with number of quantum n=1, such as scattering, photoeffect or fluorescence, an enhancement factor of three to four orders of magnitude is required. But for nonlinear processes of ionization and the hight order harmonic generation the required enhancement factor is near 50. Plasmon polariton superfocusing can be used for spatially localized coherent control excitons in a semiconductor quantum dot. In such experiments, short laser $\pi$-pulses are used to coherently manipulate the exciton state.

A pump pulse prepares an exciton in a well defined exited state. To realize a quantum gate with the help of couple nearby quantum dots it is necessary to control resonant energy transfer between them. Energy levels tuning can be obtained utilizing the Stark effect in a nonresonant field of terahertz pulses. Control the exciton dynamics in a dot array makes quantum dot solids ideal for us as nanostructures for nanoscale computing and quantum information process.

## Acknowledgements

The research was supported by the Russian Foundation for Basic Research (grant No. 13-07-0027, grant No. 015-07-09123 A) and by the Government order of the RF Ministry of Education and Science (project No. 1940).